# "Hot" 2DHG states in Tellurium

Shu-Juan Zhang[1], Lei Chen[2], Shuang-Shuang Li[1], Zhao-Cai Wang[1], Ying Zhang[1], Jing-Shi Ying[1], Julie Karel[3], Weiyao Zhao[3], and Ren-Kui Zheng[1]

[1]School of Physics and Materials Science, Nanchang University, Nanchang 330031, China

[2]School of Physics and Materials Science, Guangzhou University, Guangzhou 510006, China

[3]Department of Materials Science & Engineering, & ARC Centre of Excellence in Future Low-Energy Electronics Technologies, Monash University, Clayton VIC 3800, Australia

Element semiconductor Te is very popular in both fundamental electronic structure study, and device fabrication research area due to its unique band structure. Specifically, in low temperatures, Te possesses strong quantum oscillations with magnetic field applied in basal plane, either following Shubnikov-de Haas (SdH) oscillation rule or following log-periodic oscillation rule. With magnetic field applied along the [001] direction, the SdH oscillations are attributed to the two-dimensional hole gas (2DHG) surface states. Here we reported an interesting SdH oscillation in Te-based single crystals, with the magnetic field applied along the [001] direction of the crystals, showing the maximum oscillation intensity at ~ 75 K, and still traceable at 200 K, which indicates a rather "hot" 2DHG state. The nontrivial Berry phase can be also obtained from the oscillations, implying the contribution from topological states. More importantly, the high temperature SdH oscillation phenomena are observed in different Te single crystals samples, and Te single crystals with nonmagnetic/magnetic dopants, showing robustness to bulk defects. Therefore, the oscillation may be contributed by the bulk symmetry protected *hot* 2DHG states, which will offer a new platform for high-temperature quantum transport studies.

Tellurium is an elemental Weyl semiconductor with fascinating quantum transport phenomena, e.g., quantum Hall effect with nontrivial Berry phase [1], log-scale quantum oscillation [2], chiral magnetoresistance [3] etc. Te has a characteristic helical crystal structure, containing three atoms in a cell that are arranged in a hexagonal array. The helical Te chains can be either right- or left-handed arranged, which corresponds to two enantiomorphic chiral space groups, $P3_121$ and $P3_232$, as verified by scanning transmission electron microscopy [3]. Another interesting effect is related to the shape chirality of as-grown nanocrystalline Te, which suggests that the screw dislocation plays more important role than the chiral structure of Te chain [4]. The proved chiral crystal symmetry in Te provides an ideal platform to study the chirality related electronics, e.g., Weyl fermions and chiral magnetoresistance, as well as, paves the way for the development of chirality-based devices [3, 5].

The broken inversion symmetry in Te hosts Weyl nodes in its electronic band structure, which is first predicted by Hirayama *et al.*[6] in 2015. At ambience pressure, Te has a direct band gap of ~ 0.314 eV, which closes upon application of external pressure, e.g., forming a Weyl semimetal state at ~ 2.2 GPa [6]. The band structure is verified by spin and angular resolved photoemission spectroscopy (spin-ARPES) [7], in which the authors confirmed the band gap of ~ 330 meV, multiple Weyl nodes, and radial spin texture. Due to the pining effect of Te vacancies, the Fermi level of as-grown Te single crystals is usually at the valence band edge, yielding metallic temperature dependent transport behaviors at high temperatures, e.g., above ~ 30 K, below which the resistivity increases with cooling, showing an insulating-like behavior [2]. Under external pressure, Te shows gap closing in magnetotransport study, and exhibits semimetal to Weyl semimetal transition feature at ~2 Pa [8], agreeing well with the prediction [6]. At low temperatures, Te shows clear resistivity oscillation effects with external magnetic fields, which, however, follows different rules. Zhang *et al.*[2] observed log-scale oscillations, indicating discrete scale invariance bounded states [9] beyond quantum limits;

Akiba *et al*.[10] observed traditional Shubnikov-de Haas (SdH) oscillations instead. Note that, in their experiments, the direction of the magnetic fields is perpendicular to the [001] direction. With magnetic field along the [001] direction, K. von Klitzing et al., observed SdH oscillations in the second derivative magnetoresistance (MR) curve, which was attributed to the 2DHG states on [001] surfaces.[11]

Due to the special crystal structure and electronic states in Te, the attempt of its applications in electronic device never stops since its topology was predicted. Thin-Te-layer based field effect transistors (FET) have been fabricated to demonstrate its advantages in electronics: e.g., high mobility and air stability of thin films/flakes [12-15], quantum Hall effect with nontrivial Berry phase at low temperatures [1, 16], great current-carrying ability of Te chains encapsulated in nanotubes [17], and low-OFF current density in phase change memory devices [18].

Here, we focused on the quantum oscillations on [001] direction, which is not reported in recent single-crystal quantum transport studies [2, 8, 10, 19]. It is worth to note that, even in Klitzing's experiments[11], the oscillation is too weak to be notice in the original MR curve at 2 K. However, when we heated the sample to a relatively high temperature, strong SdH oscillations with nontrivial Berry phase emerged, indicating a "hot" 2DHG state. Moreover, the "hot" 2DHG state is quite robust against bulk defects, nonmagnetic and magnetic dopants, which can be potentially used in electronics.

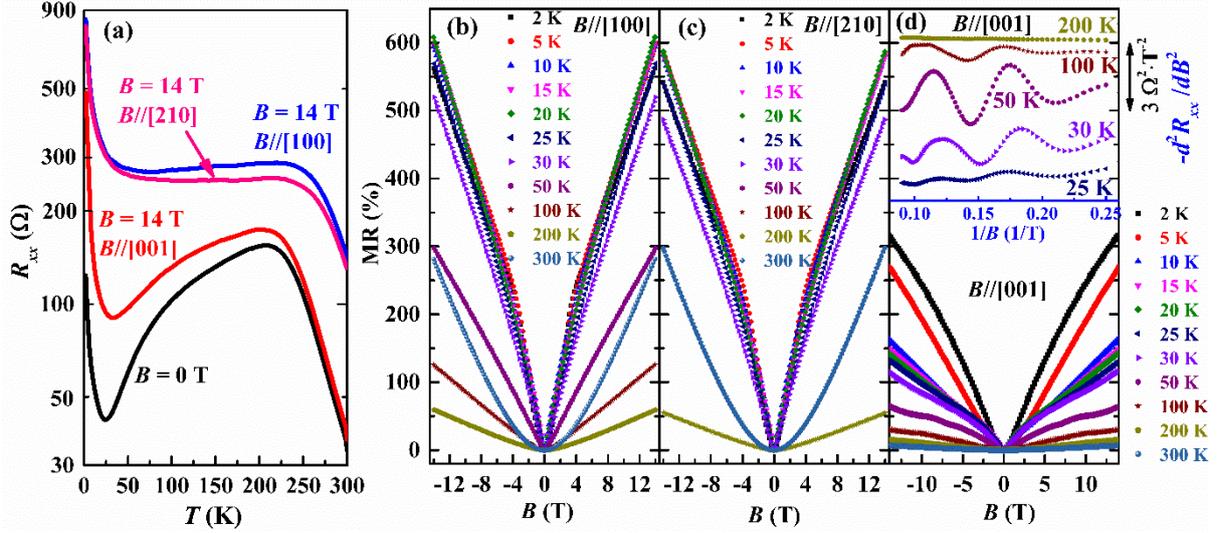

**Fig. 1** The transport properties of Te single crystal. (a) The temperature dependent RT curves, with $B = 0$, and $B = 14$ T. Note that, the magnetic fields are applied in three configurations: along (100), (210) and (001) direction, respectively. (b-d) The MR curves at various temperatures from 2 K to 300 K, with magnetic field along different axes. The Inset of Panel (d) show the SdH oscillation patterns obtained from Panel (d).

Fig. 1 shows the electronic transport properties of a pure Te single crystal in the 2 – 300 K temperature region and magnetic fields up to 14 T. Under zero magnetic field, the resistance shows an insulating behavior upon cooling from 300 to 220 K and metallic behavior upon further cooling to 25 K [black curve in Fig. 1(a)]. This insulating-to-metallic behavior is significantly different from that report in Ref. [2], where the authors only observed metallic behavior of the resistance upon cooling from 300 K. We fit the resistance data in insulating region using the Arrhenius equation $\ln\rho = \ln\rho_0 + E_g/2k_BT$, where $E_g$ is the transport band gap, $k_B$ is the Boltzmann constant, and $T$ is the temperature, as shown in Supplementary Fig. S1. The gap shown here is 205±10 meV, which is smaller than the theoretical prediction [6] and ARPES results [7], probably due to the Te vacancy defect level. Upon the application of a 14 T magnetic field, the resistance increases considerably, regardless of the direction of the magnetic field is along the [001], [100] or [210] direction. However, the MR for $B//[001]$ is much weaker than that for $B//[100]$ and $B//[210]$, demonstrating significant anisotropy upon rotating the direction of the magnetic field from the [001] direction to the basal plane. The MR anisotropy within the basal plane [schematically illustrated in Fig. 1(a)], however, is rather weak.

Fig. 1(b) shows the MR vs. *B* curves, as measured at certain temperatures, with the magnetic fields applied along the [100] direction. Below 100 K, MR increases roughly linearly with the magnetic field. This linear MR (LMR) can be interpreted from two aspects: the mobility fluctuation due to defects [20], and the quantum LMR, which describes the linear increase of resistance with magnetic field above the quantum limit [21]. In the mobility fluctuation model, the carrier's rajectories in a large enough magnetic field can be described in the combination of fast cyclotron orbit motion *r*(*t*) and slow guiding-center (GC) motion *R*(*t*). The cyclotron motion *r*(*t*) with the characteristic radius $r_c$ depends on materials' fermiology and magnetic field, whereas the GC motion *R*(*t*) depends on the potential of electrons over one cycle [22]. In a material with slowly varying disorder potentials (the potential disorder correlation length $\xi \gg r_c$), electron trajectories are dominated by the GC motion, which follows the local disorder landscape. In this case, MR is dominated by the GC motion and thus shows linear dependence on the magnetic field above the turn-on magnetic field [23]. At ~ 4.5 T, a bump occurs on the MR curve, which is the quantum limit of Te single crystal, The MR vs. *B* curve displays a weak bump near *B*=4.5 T which is the quantum limit of Te single crystal. Above 4.5 T, the variation of the resistance with the magnetic field follows the quantum LMR model. In Ref. 2, the Te single crystal shows the log-scale quantum oscillations above 4.5 T, which is absence here. With magnetic fields rotate from the [100] direction, the LMR effect is still observed (Supplementary Fig. 3).

Fig. 1(c) shows the MR vs. *B* curves with the direction of the magnetic field along the [210] direction. Note that, the [100] and [210] directions are two perpendicular basal-plane vectors in Te single crystals. Since the MR anisotropy is quite weak within the basal plane, the MR behaviors are almost the same as those shown in Fig. 1(b). However, the MR behaviors for *B*//[001] [Fig. 1(d), lower panel] are quite different from those for *B*//[100] and *B*//[210].

For $B//[001]$, LMR dominants the MR behavior below ~15 K, and is significant smaller than that for $B//[100]$ and $B//[210]$, e.g., at 2 K and 14 T, the MR is ~600% with the magnetic field applied in basal plane, and ~300 % with the magnetic field applied along the [001] direction. Another important point is the clear sign of periodic oscillation of the resistance at 50 K. To clearly demonstrate the oscillation patterns, we calculated the second derivative of the resistance and show them in the upper panel of Fig. 1(d). The resistance shows periodic oscillation with $1/B$ scale, a signature of SdH oscillation. Usually, the amplitude of SdH oscillation damps with heating, which is a function of effective electron mass of carriers contributing to the oscillation. Unexpectedly, the oscillation is still very strong at 50 K, and its amplitude damps with both cooling and heating, e.g., the oscillation is weak at 25 K, and negligible at 200 K. To study this anomalous temperature dependent SdH oscillation effect, we further conducted electronic transport measurements on Zr:Te crystals.

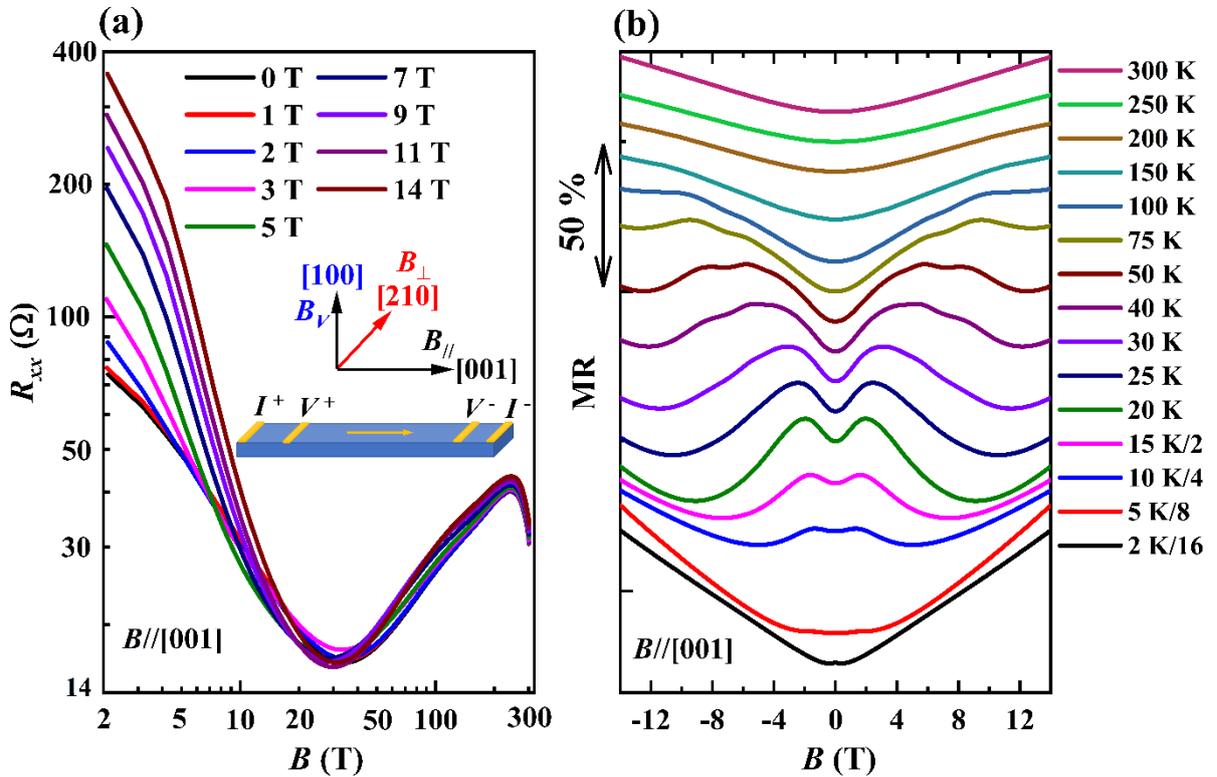

**Fig. 2** Magnetotransport properties of $Zr_{0.03}Te_{0.97}$ single crystals. (a) Temperature dependence of the resistance by applying magnetic fields along the [001] direction. (b) The MR as a function of the magnetic field at various temperatures for $B//[001]$. Note that the low temperature curves are shown in scale-down mode.

As a representative, the magnetotransport properties of a 3 at% Zr-doped Te single crystal for $B//[001]$ is shown in Fig. 2. Note that, the magnetotransport properties of this crystal for $B//[100]$ and $B//[210]$ are shown in Supplementary Fig. S4 and S5, respectively. For the $Zr_{0.03}Te_{0.07}$ crystal, the resistance shows insulator-metal-insulator transition with heating from 2 to 300 K. The high temperature metal-to-insulator transition occurs at $T_{MI}$~242 K, which is 34 K higher than that ($T_{MI}$~208 K) of the Te crystal [Fig. 1(a)], which indicates that the Zr:Te crystals are more metallic than the Te crystal. As aforementioned, one of the advantages of Te-based electronic is its capacity of electric current. Therefore, Zr doping could be a good approach to further increase the conductivity of Te-based electronics. Instead of showing positive MR in the 2 ~ 300 K temperature region for $B//[001]$ for the Te crystals, the Zr:Te crystal shows negative MR in certain temperature region (approximately 10 ~ 75 K) by applying magnetic fields along the [001] direction. The negative MR in Zr:Te is more clearly demonstrated by the MR vs. $B$ curves shown in Fig. 2(b), in which the low temperature curves are shown in scale-down plotting, e.g., the 2-K curve has been divided by 16. The positive MR dominating total behavior below ~ 10 K and above ~ 100 K, between which the crystal exhibits negative MR effect at certain magnetic fields. As discussed above, the crystal symmetry remains unchanged with Zr doping. Therefore, the chiral anomaly could be the reason to the negative MR in the present nonmagnetic Zr doped Te single crystals. Similar negative MR was previously reported in Te single crystals in Ref. 2, in which the authors show that the negative MR originates from the chiral anomaly. Besides the negative MR, the resistance shows clear SdH oscillations between 30 and 150 K, which will be discussed in Fig. 3.

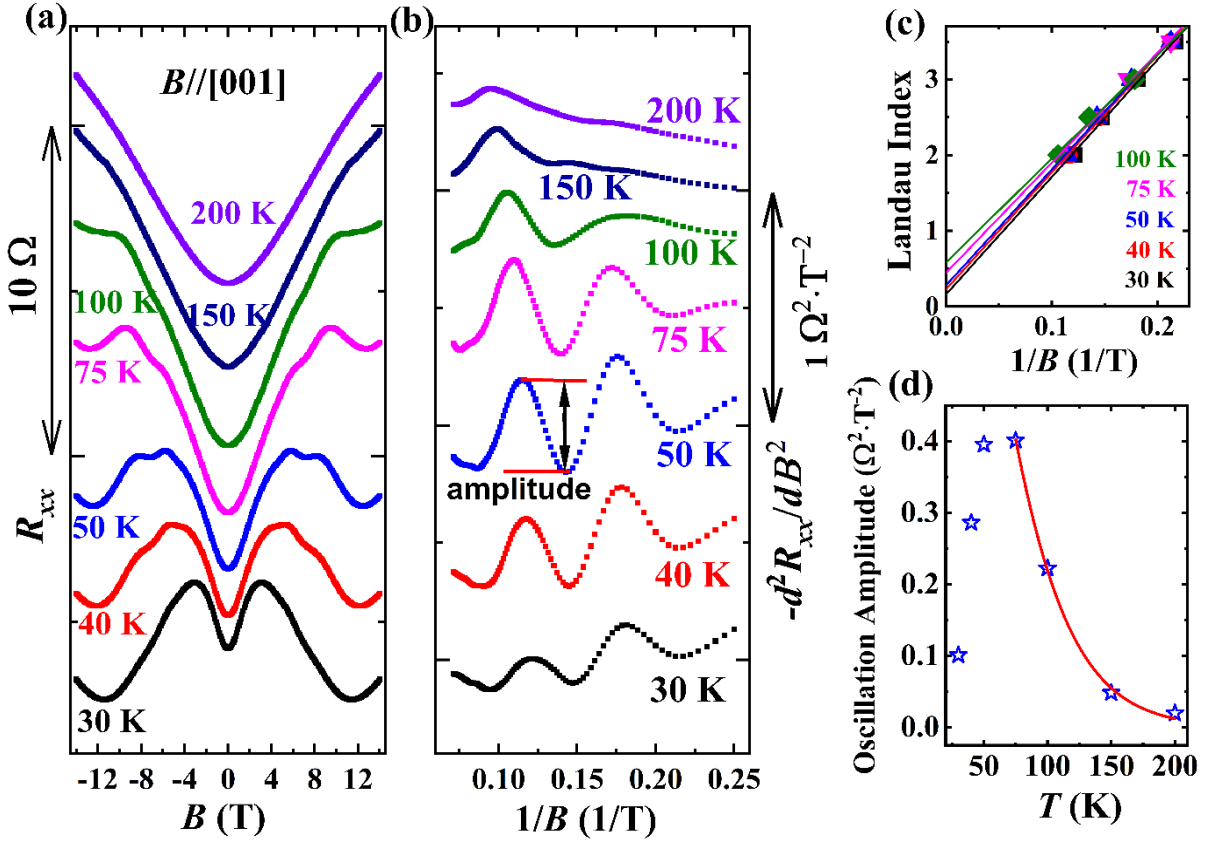

**Fig. 3** The SdH oscillation of $Zr_{0.03}Te_{0.07}$ single crystals. (a) The resistance as a function of magnetic field applied along the [001] direction. (b) The SdH oscillation patterns obtained by calculating the second derivative of the resistance shown in (a), which are plotted against $1/B$. (c) The Landau fan diagram at different temperatures. (c) The oscillation amplitude as a function of temperatures. The red line is the fitting to the data using the LK formula.

To quantify the SdH oscillations, we show the raw oscillation patterns (i.e., the resistance vs. $B$ data) of $Zr_{0.03}Te_{0.97}$ crystal in Fig. 3(a) to avoid the influences of the normalization of the resistance on the oscillation amplitude. The shapes of the background of the resistance vs. $B$ curves are quite complicate for the low-frequency oscillations. The second derivative calculation method is employed here to isolate the cosine oscillatory term from the polynomial-shape backgrounds. As shown in Fig. 3(b), the quantum oscillations follow the SdH rule, which is evidenced by the periodicity on $1/B$ axis. The SdH oscillations of $Zr_{0.03}Te_{0.97}$ show the same anomalous temperature dependent behaviors as those of Te crystals [upper panel of Fig. 1(d)]. Namely, the oscillation amplitude increases with increasing temperature from 30 to 75 K and decreases with further increase in the temperature. Notably, there still exists oscillation signal up to 200 K. Another unusual feature of these oscillations is the frequency shifting with temperature. To better understand the fermiology along the [001] direction, we employ the Lifshitz-Kosevich (LK) equation to describe the oscillation patterns:

$$\frac{\Delta\rho}{\rho(0)} = \frac{5}{2}(\frac{B}{2F})^{\frac{1}{2}} R_T R_D R_S \cos\left(2\pi\left(\frac{F}{B}+\gamma-\delta\right)\right)$$

where $R_T = \alpha T m^*/B\sinh(\alpha T m^*/B)$, $R_D = \exp(-\alpha T_D m^*/B)$, and $R_S = \cos(\alpha g m^*/2)$. Here, $m^*$ is the ratio of the effective cyclotron mass to the free electron mass $m_e$; $g$ is the $g$-factor; $T_D$ is the Dingle temperature; and $\alpha = (2\pi^2 k_B m_e)/\hbar e$, where $k_B$ is Boltzmann constant, $\hbar$ is the reduced Planck constant, and $e$ is the elementary charge. The oscillation of $\Delta\rho$ is described by the cosine term with a phase factor $\gamma - \delta$, in which $\delta = 0$ for 2D systems, and $\pm 1/8$ for 3D Fermi pockets, $\gamma = 1/2 - \Phi_B/2\pi$, where $\Phi_B$ is the Berry phase. In electron systems, the Berry phase is usually 0. According to the Onsager-Lifshitz equation, the frequency of quantum oscillations, $F = (\varphi_0/2\pi^2) A_S$, where $A_S$ is the extremal cross-sectional area of the Fermi surface perpendicular to the direction of magnetic field, and $\varphi_0$ is the magnetic flux quantum. The temperature and magnetic damping factor $R_T$ and $R_D$ are related to the cyclotron mass and quantum relaxion time of electron systems. The Berry phase can be obtained by Landau fan diagram, as shown in Fig. 3(c). The linear fitting curves of the oscillation peaks/valleys with integer/half integer show non-zero intercept values, e.g., between 0.15 and 0.58, which is the evidence of the nontrivial Berry phase of the SdH oscillations.

The effective electron mass is tricky here, due to the frequency shifting and anomalous temperature dependence of the oscillation, as discussed above. To estimate this important parameter, we choose the derivative resistance difference between last oscillation peak in high magnetic fields and the next valley on lower field side as "oscillation amplitude" [Fig. 3(b)], which are summarized in Fig. 3(d). A clear damping with temperature is observed from 75 to 200 K, which can be fitted with the LK formula to estimate the $m^* \sim 0.02$. The ultralight fermions with nontrivial Berry phase in such a system is probably the Weyl fermions. We note that similar SdH oscillations are also observed in the 5% and 7% Zr doped Zr:Te single crystals, as shown in Supplementary Fig. S6.

For the 5% doped crystals, the 40 – 150 K oscillations are similar with the observation in the 3% Zr doped Zr:Te crystals. In 7% Zr doped crystal, the quantum oscillations are also observed below ~ 10 K. The strong temperature dependent damping behavior suggests large effective electron mass. At high temperatures, the SdH oscillations show the similar temperature dependence behaviors with the Te crystals. Therefore, we deduce that, the SdH oscillations below 10 K and at high temperatures are contributed by different Fermi pockets.

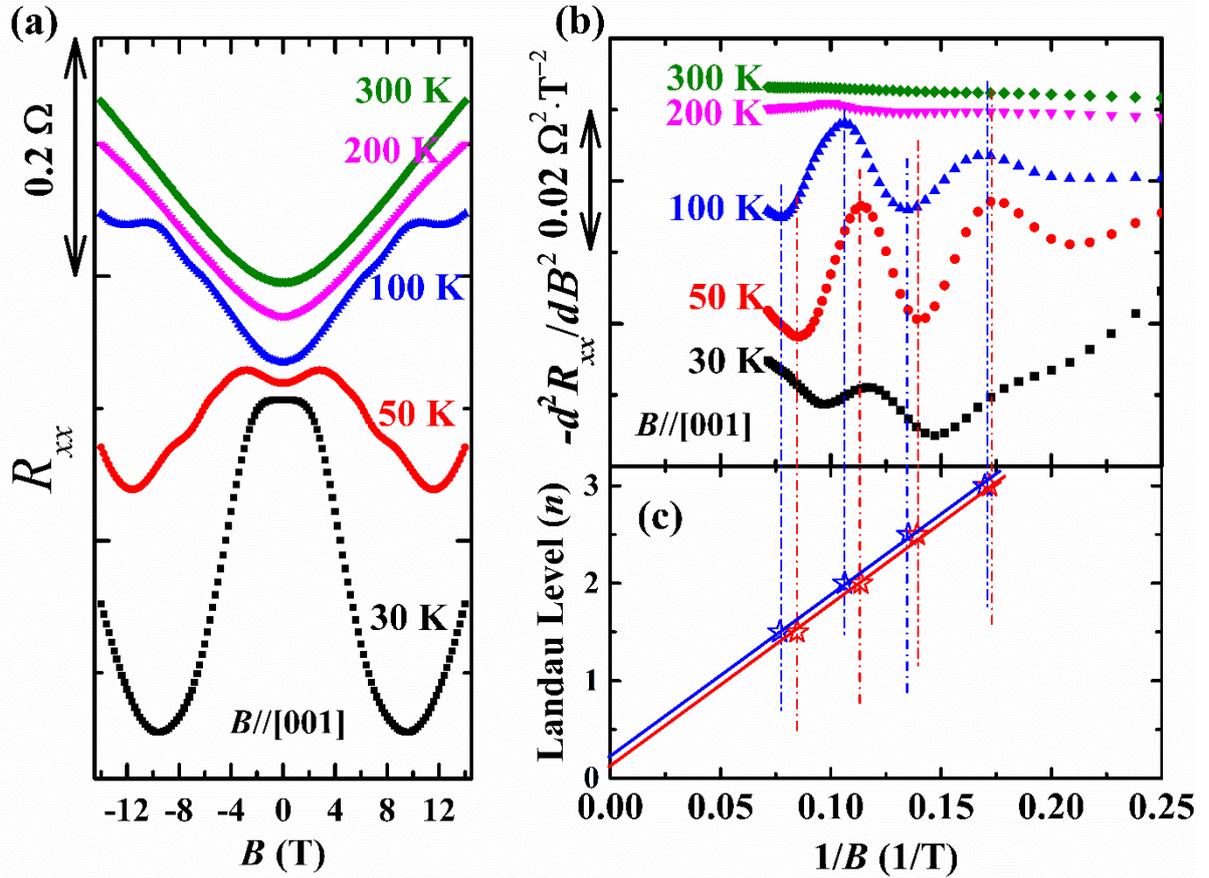

**Fig. 4** The SdH oscillation of Cr:Te single crystals. (a) The resistance with magnetic fields applied along (001) direction at various temperatures. (b) The oscillation patterns obtained from the MR curves in Panel (a). (c) The Landau fan diagram related to SdH oscillations at 50 and 100 K.

To further verify whether this anomalous temperature dependent SdH oscillations occur in Te-based single crystals with magnetic dopants, we revisit the MR behaviors of our Cr:Te single crystals whose low-temperature ($T$<30 K) MR behaviors has been discussed. Here, we focus on the high temperature oscillations, as shown in Fig. 4. Upon heating from 30 K, the negative MR decreases and changes to positive MR for $T \geq 100$ K, which is similar to the variation tendency of Zr:Te crystals. Therefore, we argue this negative MR is not contributed by the magnetic ordering of Cr dopants. As shown in Fig. 4(a), the SdH oscillation signal can be clearly observed in the Cr:Te crystals with the direction of the magnetic field along the [001] direction. We subtracted the backgrounds of MR to extract the oscillation patterns, which are plotted against in $1/B$ in Fig. 4(b). One should notice that the variation tendency of the amplitudes of the oscillations with temperature is also similar to those of Te and Zr:Te crystals, and moreover, the SdH oscillations also possess nontrivial Berry phase, which can be verified by the Landau fan diagram shown in Fig. 4(c).

Such massless topological fermions contributed quantum oscillations in Te and Zr/Cr doped single crystals are quite robust against bulk defects, nonmagnetic and magnetic dopants, which is very similar to the 3D topological insulators. Therefore, the massless fermions contributing to the SdH oscillations these single crystals could be due to the Fermi arc surface state. Quantum transport study related to the Fermi arc is rarely reported because the metallic bulk states usually dominate the electronic transport behaviors. For semiconductor-based Te Weyl system, the 2DHG dominated SdH oscillation behavior manifests even up to ~200 K is quite interesting and may attract further study of the materials system. Another interesting part is the temperature dependent oscillation frequencies, which indicates that the evolution of the Fermi pocket size with heating. This point can be further verified by ARPES study.

In summary, we reported the magnetotransport properties of high-quality Te, nonmagnetic Zr-doped and magnetic Cr-doped Te single crystals, which all shows the "hot" 2DHG state dominated SdH oscillations. Taking the $Zr_{0.03}Te_{0.97}$ as an example, the amplitude of the oscillations increases with heating from 30 to 75 K and damps with further heating, and still traceable at a high temperature of 200 K with 14 T magnetic field. A fitting of the oscillation amplitude within the high temperature region (75–200 K) using the LK formula yields $m^* \sim 0.02$, which suggests massless feature of the 2DHG. The analysis of the SdH oscillations provides evidence for a nontrivial Berry phase and robustness of the oscillation against bulk defects, nonmagnetic and magnetic dopants. The ultralight effective electron mass with nontrivial Berry phase of 2DHG state indicating its topological origination (probably Weyl states). The hot 2DHG state in tellurium offers a high temperature quantum transport platform for fundamental studies and electronic device applications.

**Acknowledgements**

This work was supported by the National Natural Science Foundation of China (Grant No. 11974155). JK and WZ acknowledge Australian Research Council Discovery Project DP200102477 and ARC FLEET Project.

**References**

[1] G. Qiu, C. Niu, Y. X. Wang, M. W. Si, Z. C. Zhang, W. Z. Wu, and P. D. D. Ye, Quantum Hall effect of Weyl fermions in n-type semiconducting tellurene, Nat. Nanotechnolo. **15**, 585 (2020).

[2] N. Zhang, G. Zhao, L. Li, P. Wang, L. Xie, B. Cheng, H. Li, Z. Lin, C. Xi, and J. Ke, Magnetotransport signatures of Weyl physics and discrete scale invariance in the elemental semiconductor tellurium, Proc. Natl. Acad. Sci. U.S.A. **117**, 11337 (2020).

[3] F. Calavalle, M. Suárez-Rodríguez, B. Martín-García, A. Johansson, D. C. Vaz, H. Yang, I. V. Maznichenko, S. Ostanin, A. Mateo-Alonso, A. Chuvilin, I. Mertig, M. Gobbi, F. Casanova, and L. E. Hueso, Gate-tuneable and chirality-dependent charge-to-spin conversion in tellurium nanowires, Nat. Mater. s41563 (2022).

[4] A. Ben-Moshe, A. Da Silva, A. Müller, A. Abu-Odeh, P. Harrison, J. Waelder, F. Niroui, C. Ophus, A. M. Minor, and M. Asta, The chain of chirality transfer in tellurium nanocrystals, Science **372**, 729 (2021).

[5] T. Furukawa, Y. Watanabe, N. Ogasawara, K. Kobayashi, and T. Itou, Current-induced magnetization caused by crystal chirality in nonmagnetic elemental tellurium, Phys. Rev. Res. **3**, 023111(2021).

[6] M. Hirayama, R. Okugawa, S. Ishibashi, S. Murakami, and T. Miyake, Weyl node and spin texture in trigonal tellurium and selenium, Phys. Rev. Lett. **114**, 206401 (2015).

[7] G. Gatti, D. Gosálbez-Martínez, S. Tsirkin, M. Fanciulli, M. Puppin, S. Polishchuk, S. Moser, L. Testa, E. Martino, and S. Roth, Radial Spin Texture of the Weyl Fermions in Chiral Tellurium, Phys. Rev. Lett. **125**, 216402 (2020).

[8] T. Ideue, M. Hirayama, H. Taiko, T. Takahashi, M. Murase, T. Miyake, S. Murakami, T. Sasagawa, and Y. Iwasa, Pressure-induced topological phase transition in noncentrosymmetric elemental tellurium, Proc. Natl. Acad. Sci. U.S.A. **116**, 25530 (2019).

[9] H. Wang, H. Liu, Y. Li, Y. Liu, J. Wang, J. Liu, J. Y. Dai, Y. Wang, L. Li, and J. Yan, Discovery of log-periodic oscillations in ultraquantum topological materials, Sci. adv. **4**, eaau5096 (2018).

[10] K. Akiba, K. Kobayashi, T.C. Kobayashi, R. Koezuka, A. Miyake, J. Gouchi, Y. Uwatoko, and M. Tokunaga, Magnetotransport properties of tellurium under extreme conditions, Phys. Rev. B **101**, 245111 (2020).

[11] K. Von Klitzing, G. Landwehr, Surface quantum states in tellurium, Solid State Communications, 9 (1971) 2201-2205.

[12] Y. Wang, G. Qiu, R. Wang, S. Huang, Q. Wang, Y. Liu, Y. Du, W. A. Goddard, M. J. Kim, and X. Xu, Field-effect transistors made from solution-grown two-dimensional tellurene, Nat. Electron. **1**, 228 (2018).

[13] C. S. Zhao, C. L. Tan, D. H. Lien, X. H. Song, M. Amani, M. Hettick, H. Y. Y. Nyein, Z. Yuan, L. Li, M. C. Scott, and A. Javey, Evaporated tellurium thin films for p-type field-effect transistors and circuits, Nat. Nanotechnolo. **15**, 53 (2020).

[14] C. Zhao, H. Batiz, B. Yasar, H. Kim, W. Ji, M. C. Scott, D. C. Chrzan, and A. Javey, Tellurium Single-Crystal Arrays by Low-Temperature Evaporation and Crystallization, Advanced Materials **33**, 2100860 (2021).

[15] M. Peng, R. Xie, Z. Wang, P. Wang, F. Wang, H. Ge, Y. Wang, F. Zhong, P. Wu, and J. Ye, Blackbody-sensitive room-temperature infrared photodetectors based on low-dimensional tellurium grown by chemical vapor deposition, Sci. Adv. **7**, eabf7358 (2021).

[16] C. Niu, G. Qiu, Y. X. Wang, M. W. Si, W. Z. Wu, and P. D. D. Ye, Bilayer Quantum Hall States in an n-Type Wide Tellurium Quantum Well, Nano Lett. **21**,7527 (2021).

[17] J. K. Qin, P. Y. Liao, M. Si, S. Gao, G. Qiu, J. Jian, Q. Wang, S. Q. Zhang, S. Huang, and A. Charnas, Raman response and transport properties of tellurium atomic chains encapsulated in nanotubes, Nat. electron. **3**, 141 (2020).

[18] J. Shen, S. Jia, N. Shi, Q. Ge, T. Gotoh, S. Lv, Q. Liu, R. Dronskowski, S. R. Elliott, and Z. Song, Elemental electrical switch enabling phase segregation–free operation, Science **374**,1390 (2021).

[18] S. Lin, W. Li, Z. Chen, J. Shen, B. Ge, and Y. Pei, Tellurium as a high-performance elemental thermoelectric, Nat. commun. **7**, 10287 (2016).

[19] J. F. Oliveira, M. B. Fontes, M. Moutinho, S. E. Rowley, E. Baggio-Saitovitch, M. S. B. Neto, and C. Enderlein, Pressure-induced Anderson-Mott transition in elemental tellurium, Commun. Mater. **2**, (2021).

[20] Y. Du, G. Qiu, Y. Wang, M. Si, X. Xu, W. Wu, and P. D. Ye, One-dimensional van der Waals material tellurium: Raman spectroscopy under strain and magneto-transport, Nano

lett. **17**, 3965 (2017).

[21] J. Feng, Y. Pang, D. Wu, Z. Wang, H. Weng, J. Li, X. Dai, Z. Fang, Y. Shi, and L. Lu, Large linear magnetoresistance in Dirac semimetal $Cd_3As_2$ with Fermi surfaces close to the Dirac points, Phys. Rev B **92**, 081306 (2015).

[22] A. Abrikosov, Quantum magnetoresistance, Phys. Rev. B **58**, 2788 (1998).

[23] J. C. Song, G. Refael, and P. A. Lee, Linear magnetoresistance in metals: Guiding center diffusion in a smooth random potential, Phys. Rev. B **92**, 180204 (2015).

[24] X. Wang, Y. Du, S. Dou, and C. Zhang, Room temperature giant and linear magnetoresistance in topological insulator $Bi_2Te_3$ nanosheets, Phys. Rev. Lett. **108**, 266806 (2012).

**Supplementary Materials**

High-quality Te, Zr:Te and Cr:Te single crystals were grown via the PVD method. In detail, 4N purity Cr, Zr, and Te powders with element doping level of 1%, 3%, 5%, 7% were sealed in a silica tube in vacuum ($2\times10^{-3}$ Pa). The sealed mixture was placed in a two-temperature-zone furnace, followed by: 1) heating to 1000 °C at 1 °C/min, 2) keeping at 1000 °C for 1 hour, 3) cooling down to 400 °C and 300 °C. respectively, 4) keeping the 100 °C temperature gradient for 1 week. After the growth process, needle-like single crystals with a size of ~ 5-10×1×1 mm$^3$ were obtained.

X-ray diffraction (XRD) patterns were collected using a Rigaku SmartLab X-ray diffractometer equipped with Cu Kα1 radiation (λ=1.5406 Å). Raman spectra were recorded on a Raman spectrometer equipped with a 532 nm laser (Witec Alpha300R) as an excitation source at room temperature. The magnetotransport properties were measured using a physical property measurement system (PPMS, DynaCool-14, Quantum Design). Standard four-probe configuration was fabricated on the shiny surface of a single crystal using indium. For magnetotransport measurements, the electric current is always along the [001] direction of crystals and the direction of the magnetic field is applied along either perpendicular or parallel to the electric current.

Fig. S1(a) shows the room-temperature Raman spectra taken on the (100) surface of as-grown single crystals. The three first-order lattice vibration modes are $E_1$, $A_1$ and $E_2$, located at 92 cm$^{-1}$, 123 cm$^{-1}$ and 140 cm$^{-1}$, respectively, consistent with previously reports [17, 24]. The $A_1$ mode describes chain expansion, where atoms move in the basal plane. The $E_1$ and $E_2$ modes represent bond bending and stretching, respectively. Fig. 1(b) demonstrates the room-temperature XRD θ-2θ patterns of Zr doped Te single crystals. The comb-like (*h*00) peaks confirm the perfect orientation of as-grown single crystals. From the XRD and Raman patterns, we conclude that the element doped Te single crystals share the same space group with Te, in which the lattice distortion is negligible. Therefore, in the following transport measurements, we believe that the doping elements only change the Fermi level and/or magnetic properties. The band structure and band topology are the same with theoretical prediction and ARPES measurements [2, 6, 7].

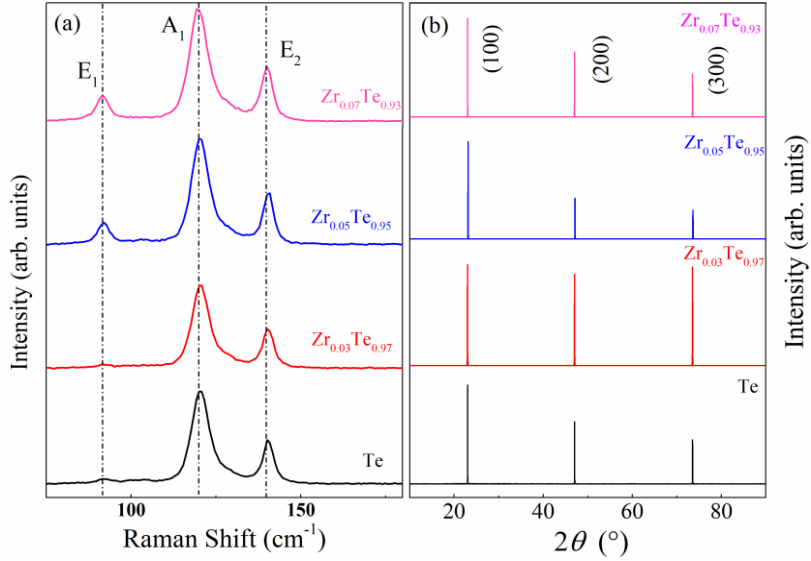

**Fig. S1.** (a) Raman spectra and (b) XRD patterns of Zr doped Te single crystals at room temperature.

A clear picture of the quantum limit phenomenon together with the angular dependent MR is demonstrated in Supplementary Fig. S3. With increase magnetic field from zero field to the "turn-on" field, the MR shows parabolic increasing trend. Above the "turn-on" field, the resistance increases linearly with magnetic field. The "bump" at ~ 4.5 T is the last peak of SdH oscillations, above which the carriers are quantized into the zeroth Landau level. The LMR effect below and above this "bump" shows different slopes with the magnetic field, which are governed by the mobility fluctuation and quantum LMR mechanisms, respectively. Note that the rotation of the sample is conducted within the basal plane, which means that the direction of the magnetic field is always perpendicular to the electric current during rotation. One may observe that the LMR has only one slope for $\theta \geq 30°$, which is dominant by the mobility flucturation mechanism.

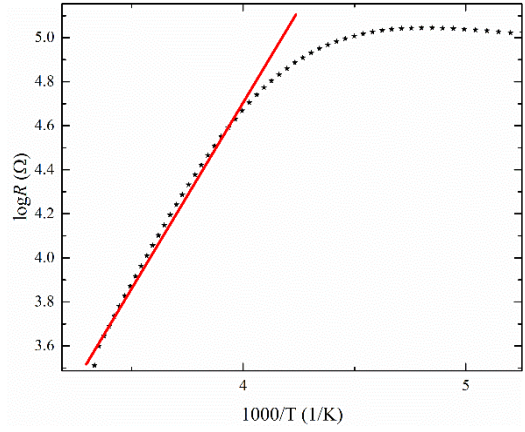

Fig. S2 The band gap fitting of Te single crystal.

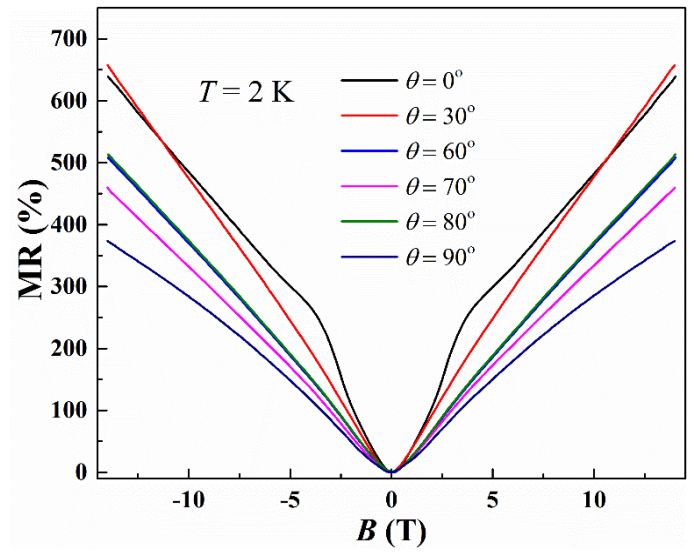

Fig. S3 The angular dependent MR curves of Te single crystal a 2 K.

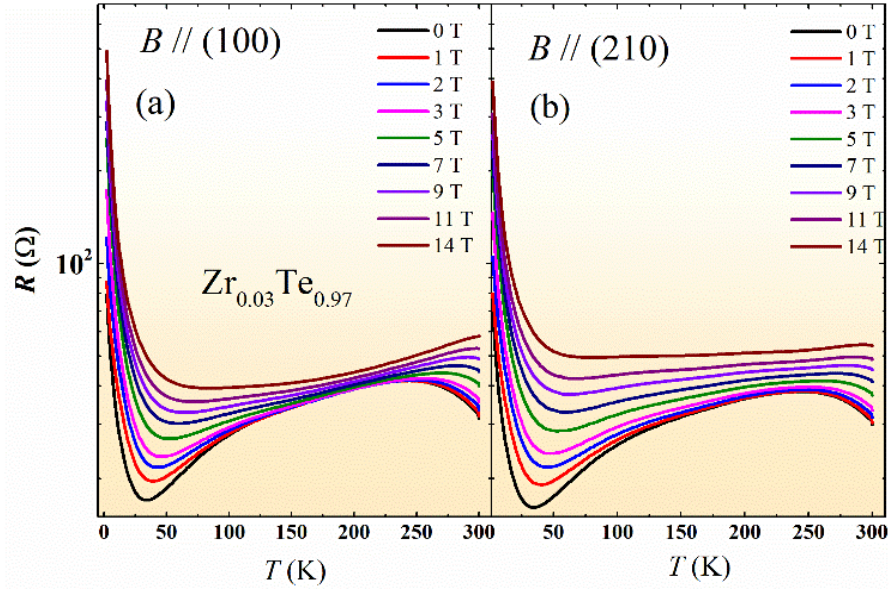

Fig. S4 The RT curves of $Zr_{0.03}Te_{0.97}$ single crystals, with magnetic field applied along: (a) the [100] direction and (b) the [210] direction.

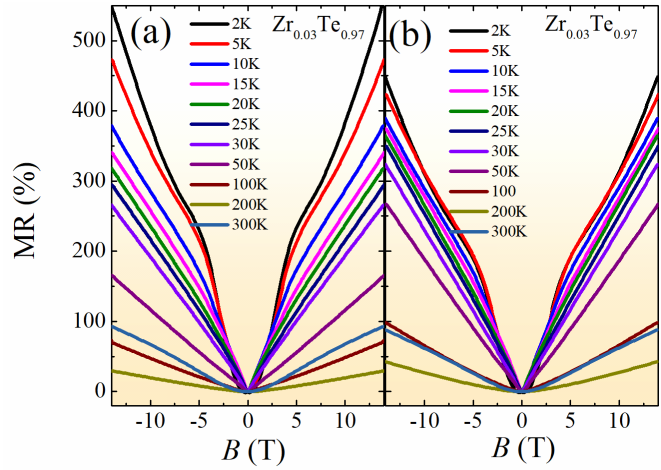

Fig. S5 The RT curves of $Zr_{0.03}Te_{0.97}$ single crystals, with magnetic field applied along: (a) the [100] direction and (b) the [210] direction.

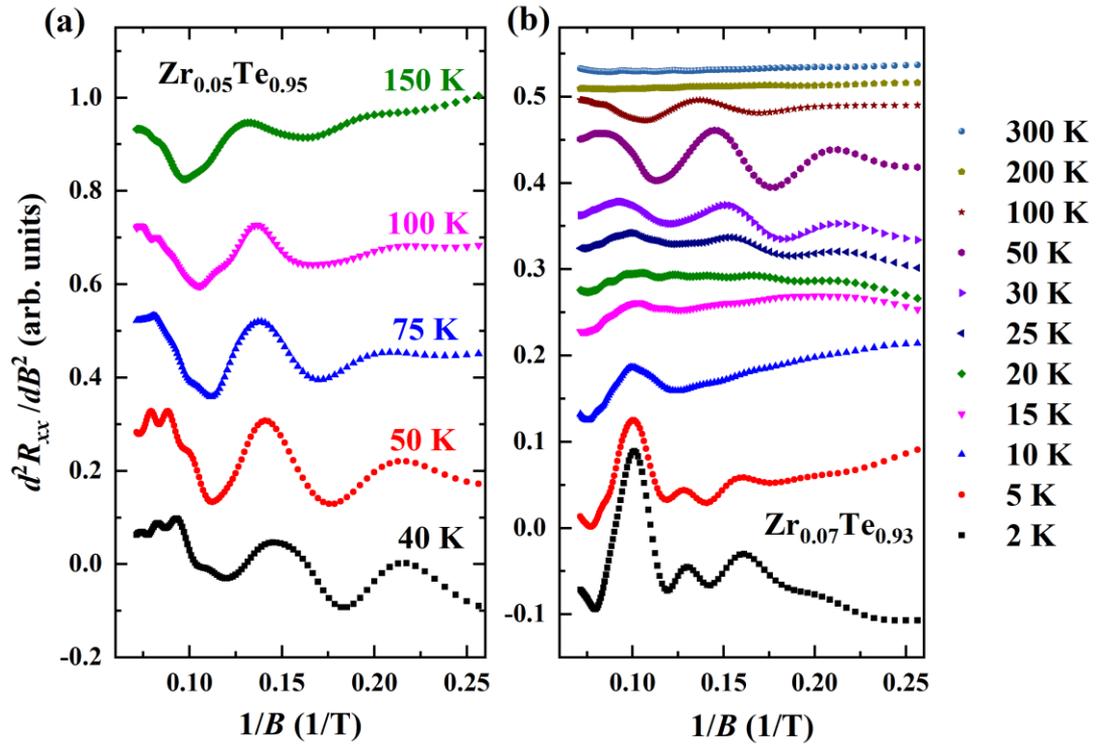

Fig. S6 The quantum oscillations in Zr0.05 and Zr0.07 doped Te single crystals, shown in Panel (a) and (b), respectively.